# High-pressure stability and compressibility of $APO_4$ (A = La, Nd, Eu, Gd, Er, and Y) orthophosphates: An x-ray diffraction study using synchrotron radiation

R. Lacomba-Perales[1,2], D. Errandonea[2,3,†] , Y. Meng[4], and M. Bettinelli[5]

[1]Malta Consolider Team, Universidad de Valencia, Edificio de Investigación, C/Dr. Moliner 50, 46100 Burjassot (Valencia), Spain

[2]Departamento de Física Aplicada-ICMUV, Universidad de Valencia, Edificio de Investigación, C/Dr. Moliner 50, 46100 Burjassot (Valencia), Spain

[3]Fundación General de la Universidad de Valencia, Edificio de Investigación, C/Dr. Moliner 50, 46100 Burjassot (Valencia), Spain

[4]HPCAT, Carnegie Institution of Washington, Building 434E, 9700 South Cass Avenue, Argonne, IL 60439, USA

[5]Laboratory of Solid State Chemistry, DB and INSTM, Università di Verona, Strada Le Grazie 15, I-37134, Verona, Italy

**Abstract:** Room temperature angle-dispersive x-ray diffraction measurements on zircon-type $YPO_4$ and $ErPO_4$, and monazite-type $GdPO_4$, $EuPO_4$, $NdPO_4$, and $LaPO_4$ were performed in a diamond-anvil cell up to 30 GPa using neon as pressure-transmitting medium. In the zircon-structured oxides we found evidence of a reversible pressure-induced structural phase transformation from zircon to a monazite-type structure. The onset of the transition is near 17 – 20 GPa. In $LaPO_4$ a non-reversible transition is found around 26 GPa, being a barite-type structure proposed for the high-pressure phase. In the other three monazites, this structure is found to be stable up to the highest pressure reached in the experiments. No additional phase transitions or evidences of chemical decomposition are found in the experiments. The equations of state and axial compressibility for the different phases are also determined. In particular, we found that in a given compound the monazite structure is less compressible than the zircon structure. This fact is attributed to the larger packing efficiency of monazite than that of zircon. The differential bond compressibility of different polyhedra is also reported and related to the anisotropic compressibility of both structures. Finally, the sequence of structural transitions and compressibilities are discussed in comparison with other orthophosphates.

[†] Corresponding author; email: daniel.errandonea@uv.es

## 1. Introduction

Orthophosphates $APO_4$ are materials that are basically composed of $PO_4$ tetrahedra and $AO_8$ or $AO_9$ (A = trivalent metal) polyhedra. They are analogous to orthosilicates, orthovanadates, and orthoarsenates. This family of oxides generally crystallizes, depending on the ionic radii of the A cation, in two different structural types: zircon (xenotime) and monazite [1], named after the natural minerals $ZrSiO_4$ and $CePO_4$. If the ionic radius of the A cation is smaller than that of Gd, the material will have the tetragonal ($I4_1/amd$, Z = 4) zircon structure. Most of the other orthophosphates have the lower-symmetry monoclinic ($P2_1/n$, Z = 4) monazite structure. These two phases exist in nature and are important accessory minerals in granitoids and rhyolites, and because of their incorporation of rare-earth elements they can effectively control the rare-earths distribution in igneous rocks [2]. In addition, the mineral xenotime ($YPO_4$) is a common accessory mineral in plutonic and metamorphic rocks. Therefore, the knowledge of the high-pressure structural behavior of orthophosphates is very relevant for mineral physics and chemistry. It is also important for petrology studies [3].

On the other hand, the members of the orthophosphate family have gained increasing attention in the last decade due to their wide potential application and interesting optical and luminescent properties [4, 5]. Furthermore, given the crystal-chemical similarity between the lanthanide and actinide elements, monazite-structured phosphates have been investigated for their use as solid-state repository for radioactive waste [6]. On top of that, orthophosphates have been proven to be promising candidates for oxidation-resistant ceramic toughening [7]. The study of the mechanical properties of orthophosphates is relevant for all these applications.

The monazite and zircon structures are closely related. Zircon can be viewed as being composed of alternating edge-sharing $AO_8$ bisdisphenoids and $PO_4$ tetrahedra forming chains parallel to the *c*-axis (see Fig. 1). In monazite, a ninth oxygen is introduced to form $AO_9$ polyhedra for the large A cation (see Fig. 1). The larger A cation causes structural distortions, specifically a rotation of the $PO_4$ tetrahedra and a lateral shift of the (100) plane, thereby reducing the symmetry from $I4_1/amd$ to $P2_1/n$.

As described above, the investigation of the mechanical properties and the compressibility of orthophosphates can provide important information for a variety of research fields. Other oxides related to the orthophosphates have been extensively studied upon compression [8]. Their compressibility has been understood and several pressure-induced structural transitions discovered. In contrast, little effort has been dedicated to the orthophosphates, most of it focused on zircon-type compounds. Zircon-structured $YbPO_4$ and $LuPO_4$ have been found to undergo phase transitions to a tetragonal scheelite-type ($I4_1/a$, Z = 4) structure at 22 and 19 GPa, respectively [9]. On the contrary, Raman spectroscopy measurements indicate that $TbPO_4$ transforms around 9.5 GPa from zircon to a lower crystal symmetry, most likely monoclinic [10]. More recently, x-ray diffraction studies were performed in $ScPO_4$ ($YPO_4$) under non-hydrostatic conditions being the zircon-scheelite (zircon-monazite-scheelite) sequence reported [11]. Regarding monazite-type orthophosphates, only a luminiscense study was performed finding that $EuPO_4$ retains the monazite structure to at least 20 GPa [12]. Finally, the compressibility of the whole series of orthophosphates has been theoretically studied by using a chemical bond theory of dielectric description [6].

Clearly, more experimental and theoretical efforts are needed in order to deepen the understanding of the properties of orthophosphates. With this aim, we have studied the structural response of monazite-type $LaPO_4$, $NdPO_4$, $EuPO_4$, and $GdPO_4$, and zircon-type $ErPO_4$ and $YPO_4$, upon compression, under nearly hydrostatic conditions, using *in situ* synchrotron x-ray diffraction. In this work, we report the occurrence of phase transitions and the bulk and axial compressibility of each oxide. The results have been compared with previous studies for a systematic understanding of the high-pressure behavior of orthophosphates.

## 2. Experimental details

Synthetic un-doped single crystals of $APO_4$ compounds (A = La, Nd, Eu, Gd, Er, and Y) were grown by spontaneous nucleation from a $PbO$-$P_2O_5$ flux (1:1 molar ratio) [13]. The reagents employed for the growths were $NH_4H_2PO_4$, PbO (both reagent grade), and $A_2O_3$ (99.99%). The batches were put in a covered Pt crucible with a tightly fitting lid and heated up to 1300 °C inside a horizontal furnace. After a soaking time of about 15 h, the temperature was lowered to 800 °C with a rate of ≈1.8 °C per hour; the crucible was then drawn out from the furnace and quickly inverted to separate the flux from the crystals grown at the bottom of the crucible. The flux was dissolved using hot diluted nitric acid. Single crystals of good optical quality were obtained. The ambient pressure tetragonal crystals sizes up to $8\times1\times0.8$ mm$^3$ and are elongated in the direction of the *c*-axis, while the ambient pressure monoclinic ones have sizes up to $3\times2\times0.8$ mm$^3$. The crystals obtained were characterized by powder x-ray diffraction and Raman spectroscopy. Single phases of zircon-type or monazite-type structure were confirmed in all samples. The refined unit-cell parameters for them were in agreement with earlier reported values [1].

Angle-dispersive powder x-ray diffraction (ADXRD) measurements were carried out at room temperature under compression up to 30 GPa using a symmetric diamond-anvil cell (DAC) at the 16-IDB station of the High Pressure Collaborative Access Team (HPCAT) - Advanced Photon Source (APS). Two experimental runs were performed for $GdPO_4$, $LaPO_4$, and $YPO_4$ and one for the rest of the studied samples. Experiments were carried out with an incident monochromatic wavelength of 0.36802 Å for $GdPO_4$, $NdPO_4$, and $LaPO_4$, of 0.37460 Å for $LaPO_4$ and $YPO_4$, of 0.40695 Å for $YPO_4$, of 0.36980 Å for $EuPO_4$, and of 0.36783 Å for $ErPO_4$. The samples used in the experiments were pre-pressed pellets prepared using a finely ground powder obtained from the as grown single crystals. These pellets were loaded in a 130-μm hole of a rhenium gasket in a DAC with diamond-culet sizes of 300-480 μm. A few ruby grains were also loaded with the sample for pressure determination [14]. Pressure was determined using the ruby scale proposed by Dewaele *et al.* [15]. Neon, which solidify near 5 GPa [16], was used as pressure-transmitting medium in order to guarantee quasi-hydrostatic conditions in the pressure range covered by the experiments [17]. The monochromatic x-ray beam was focused down to 10 x 10 $\mu m^2$ using Kickpatrick–Baez mirrors. The images were collected using a MAR345 image plate located around 350 mm away from the sample. The collected images were integrated using FIT2D [18]. The structure refinements were performed using the POWDERCELL [19] and GSAS [20] program packages.

## 3. Results and Discussion

### A. $LaPO_4$

Two experiments were performed in $LaPO_4$, one up to 13.4 GPa and the other up to 30.2 GPa. Fig. 2 shows a selection of diffraction patterns collected at different pressures. There, it can be seen that there are not noticeable changes of the diffraction patterns up to 23.8 GPa. Indeed all of them can be properly indexed considering the monazite structure. At 26.1 GPa the appearance of additional peaks can be observed, increasing their intensity upon further compression while the monazite peaks gradually lose intensity. In particular, the peaks located around $2\theta = 5^o$, $6^o$, and $8^o$, depicted by arrows in the figure, can be clearly seen at 26.1 GPa. Also extra peaks develop from 26.1 GPa up to 30 GPa. These changes in the diffraction patterns suggest the onset of a pressure-induced phase transition. The diffraction patterns collected beyond 26.1 GPa can be well explained considering the mixture of two phases, one with monazite structure and a second phase. Consequently, the pressure-driven phase transition is kinetically sluggish. No hints of decomposition of $LaPO_4$ into its component oxides were detected in the experiments.

For the high-pressure phase of $LaPO_4$ we found three candidate structures that might explain its diffraction patterns: orthorhombic barite-type structure (*Pbnm*, Z = 4), monoclinic $AgMnO_4$-type structure ($P2_1/n$, Z = 4), and monoclinic $PbCrO_4$-type structure ($P2_1/n$, Z = 4). The possibility of having any of these structures as post-monazite structures in $LaPO_4$ is quite reasonable from the crystal-chemical point of view [8]. The effects of pressure on the crystal structure of $ABO_4$ compounds can be simulated by changing the ratio of A/B cation sizes at a fixed pressure. By applying this criterion, any of the three proposed structures could be a post-monazite phase according with Ref. 8. In addition, the barite-type and $AgMnO_4$-type have been found as post-monazite structures in $CaSO_4$ [21]. The refinement of the diffraction patterns we measured beyond 26.1 GPa shows that they can be better explained by a mixture of the monazite and barite-type structures, suggesting that this one is the most possible post-monazite structure in $LaPO_4$. In particular, at 27.3 GPa we found for the orthorhombic barite-type structure the following structural parameters: $a$ = 6.463(6) Å, $b$ = 7.835(8) Å, and $c$ = 5.072(5) Å. This implies a volume collapse of about 4 % for the crystal at the proposed transition, giving indications of its first-order nature. This is a reasonable conclusion since the monazite-barite transition involves an important atomic rearrangement. In particular, the barite-type structure implies an increase of the coordination of the La cation. La is ninth-fold coordinated in monazite while it is twelve coordinated in barite. In contrast the $PO_4$ tetrahedra remain essentially unchanged in both structure types. Upon pressure release, the barite-type structure is recovered together with the monazite structure. This non-reversibility of the transition is consistent with its first-order nature.

From the refinement of our x-ray diffraction patterns we have obtained the pressure dependences of the lattice parameters for the low-pressure phase. The evolution of the structural parameters and the atomic volume (V) with pressure are shown in Figs. 3 and 4, respectively. There it can be seen that the compression of monazite-type $LaPO_4$ is anisotropic, with $a$-axis the most compressible axis. In particular, there is a slight increase of the $c/a$ axial ratio from 0.95 at ambient pressure to 0.98 at the transition pressure. This fact together with the decrease of the β angle indicate that pressure induces a gradual increase of the crystal symmetry. They are related to the fact that the $c$-axis of monazite contains edge-linked chains of $PO_4$ tetrahedra and $AO_9$ polyhedra, while the $a$-$b$ plane involves chains of $AO_9$ polyhedra which according to our results, within this plane, are more compressible than the $PO_4$ tetrahedra, resulting in the increase in *c/a* axial ratio upon compression.

The dependence of the unit-cell parameters of monazite with pressure can be fit with a linear function. These pressure dependences are given in Table I. The pressure-volume curve of Fig. 4 was analyzed using a third order Birch-Murnaghan equation of state (EOS) [21]. We determined the following EOS parameters: $V_0$ = 301.4(7) $Å^3$, $B_0$ = 144(2) GPa, and $B_0$' = 4.0(2), being these parameters the zero-pressure volume, the bulk modulus, and its pressure derivative, respectively. As can be seen in Table II, this makes $LaPO_4$ the most compressible orthophosphate among those already studied. For the high-pressure phase we found that the compression is nearly isotropic and that the bulk compressibility is similar to that of the monazite phase. In particular, by assuming $B_0$' = 4 and $V_0$ = 296.3 $Å^3$, for the barite-type phase, we obtained $B_0$ = 143(4).

**B. $GdPO_4$, $EuPO_4$, and $NdPO_4$**

In contrast to $LaPO_4$, for $GdPO_4$, $EuPO_4$, and $NdPO_4$ we did not find any evidence of either a possible pressure-induced phase transition or decomposition. For $GdPO_4$, two experiments were conducted up to 30 GPa and all the measured diffraction patterns can be assigned to the monazite structure. We obtained the same result from experiments performed on $NdPO_4$ and $EuPO_4$ up to 28 and 25 GPa, respectively. In the case of $GdPO_4$, our conclusions agree with single-crystal diffraction studies performed up to 40 GPa [23] and in the case of $EuPO_4$ with luminescence measurements carried out up to 20 GPa [12]. From the refinement of the x-ray diffraction patterns we collected, we have determined the pressure dependence of the lattice parameters for the monazite phase of the three orthophosphates. As in the case of $LaPO_4$, the compression of the crystal is anisotropic, being the *a*-axis the most compressible axis. However, the differences between axial compressibilities are not as large as in $LaPO_4$. The dependence of the different unit-cell parameters with pressure is given in Table I. Once more, the increase of the *c/a* ratio upon compression and the decrease of the β angle suggest the occurrence of a pressure-driven symmetry enhancement. We can take advantage of this typical feature of monazite-type orthophosphates to try to estimate the pressure range of stability of monazite $GdPO_4$, $EuPO_4$, and $NdPO_4$. If we assume no other phases at play in addition to monazite and the high-pressure phase of $LaPO_4$ and consider that the monazite structure becomes unstable when *c/a* becomes equal to 0.98, as it is the case for $LaPO_4$, extrapolating our results for the other studied compounds, we found transition pressures of 44, 47, and 55 GPa for $NdPO_4$, $EuPO_4$, and $GdPO_4$, respectively. Therefore, apparently the decrease of the ionic radius of the rare-earth cation favors the stability of the monazite structure.

The pressure evolutions of the atomic volume for the three compounds described in this section are given in Fig. 4. They were analyzed using a third order Birch-Murnaghan EOS [21]. We determined the following EOS parameters: $GdPO_4$, $V_0$ = 281.1(7) Å$^3$, $B_0$ = 160(2) GPa, and $B_0$’ = 3.8(2); $EuPO_4$, $V_0$ = 281.4(7) Å$^3$, $B_0$ = 159(2) GPa, and $B_0$’ = 4.3(2); and $NdPO_4$, $V_0$ = 291.1(7) Å$^3$, $B_0$ = 170(2) GPa, and $B_0$’ = 3.6(2). The agreement of the fits with the experiments is found to be good. A systematic comparison of the bulk modulus with those of other orthophosphates will be done in the last section of this work.

## C. $YPO_4$ and $ErPO_4$

Two experiments were performed on zircon-type $YPO_4$, one up to 18 GPa and the other up to 28 GPa. The experiment for $ErPO_4$ was carried out up to 28 GPa. Fig. 5 shows a selection of diffraction patterns collected at different pressures for $YPO_4$. There, it can be seen that there are not noticeable changes of the diffraction patterns up to 19.7 GPa. Up to this pressure, all the patterns can be well indexed considering the zircon structure, beyond it, extra Bragg peaks appear at 19.7 GPa, but the zircon peaks can be still identified up to 23.5 GPa. From 23.5 GPa up to 28 GPa the diffraction patterns indicate that no additional changes take place in the crystalline structure of $YPO_4$. The changes found in the diffraction patterns indicate the onset of a pressure-induced phase transition at 19.7 GPa, with the low- and high-pressure phases coexisting from this pressure up to 23.5 GPa. Similar changes were found in the diffraction patterns of $ErPO_4$. In this case, the onset of the transition was detected at 17.3 GPa and the low- and high-pressure phases coexist up to 23.3 GPa. In both compounds no additional structural transformations are found up to the highest pressure reached in the experiments and no evidence of decomposition is detected. The phase transitions are reversible with a small hystheresis.

In $YPO_4$, the diffraction patterns collected beyond 23.5 GPa can be well explained by the monazite structure. The structural parameters for this phase at 23.5 GPa are $a$ = 6.379(9) Å, $b$ = 6.448(9) Å, $c$ = 5.980(9) Å, and $\beta$ = 101.4(5)º. The transition produces a volume collapse of about 3.5 % and the phase transformation is reversible as can be seen in Fig. 5. In $ErPO_4$, the high-pressure phase is also consistent with a monazite structure; its structural parameters at 23.3 GPa are $a$ = 6.369(9) Å, $b$ = 6.397(9) Å, $c$ = 6.038(9) Å, and $\beta$ = 101.7(5)º. In this case the volume collapse at the transition is around 4.5%. In both compounds, the high-pressure monazite structure is slightly more anisotropic that the ambient pressure monazite structure of other orthophosphates.

The occurrence of the zircon-monazite transition in $YPO_4$ and $ErPO_4$ is in agreement with previous experiments done in $YPO_4$ under less hydrostatic conditions [11] with the only difference that the present transition pressure is 3 GPa higher. This fact is not rare since non-hydrostatic stresses could strongly affect transition pressures [24]. The occurrence of the zircon-monazite transition also agrees with the findings in $TbPO_4$ [6], in which the high-pressure phase should have a monoclinic structure according to Raman spectroscopy measurements [10]. However, they contrast with the presence of the zircon-scheelite transition in $LuPO_4$ and $YbPO_4$ [9] and also in $YVO_4$ [25].

The existence of different high-pressure phases is related to the different relative cation sizes in $ABO_4$ oxides [8]. Those compounds where the A/B cation size ratio is similar to that of monazites prefer to transform from zircon to monazite; and those where the A/B cation size ratio is similar to that of scheelite prefer to transform to this structure. Therefore, for those phosphates with an A-cation size near the crossover radius between zircon and monazite (e.g. Tb and Y) the zircon-monazite transition should be induced by pressure, as we found in $YPO_4$ and $ErPO_4$. Given there is an inverse relationship between pressure and temperature in $ABO_4$ compounds [26], this explanation is in full agreement with the fact that monazite is the low-temperature form of $TbPO_4$ and zircon is the high-temperature form of $GdPO_4$ [27].

From our experiments we have determined the compressibility of the unit-cell parameters of the low- and high-pressure phases of $YPO_4$ and $ErPO_4$. The results obtained for $YPO_4$ are summarized in Fig. 6. There it can be seen that in the zircon phase the *a*-axis is more compressible than the *c*-axis. As a consequence of it, the axial ratio *c/a* increases from 0.876 at ambient pressure to 0.894 near 20 GPa, approaching the axial ratio of $ZrSiO_4$ (0.906). A similar behavior has been found in $ErPO_4$ and previously for $LuPO_4$ and $YbPO_4$ [9]. Also isomorphic compounds like $YVO_4$ show the same anisotropic compressibility [25]. The origin of this behavior is related with the packing of $AO_8$ and $PO_4$ polyhedra in the zircon structure. This structure can be considered as a chain of alternating edge-sharing $PO_4$ tetrahedra and $AO_8$ dodecahedra extending parallel to the *c*-axis, with the chain joined along the *a*-axis by edge-sharing $AO_8$ dodecahedra [28]. As we will show later, in zircon phosphates the $PO_4$ tetrahedra behave basically as uncompressible units. This makes the *c*-axis less compressible than the *a*-axis as observed in our experiments. In the high-pressure monazite-type structure we find an apparent anomalous axial compression if compared with the behavior we found for low-pressure monazite phosphates. In monazite $YPO_4$ and $ErPO_4$ the compaction occurs dominantly in the *a*- and *b*-direction, while the *c*-parameter slightly increases upon compression (see Fig. 6). Also the monoclinic β angle increases with pressure. These results contrast with the pressure-induced decrease of the β angle and the reduction of the three axes (being *a* the most compressible axis) we found for monazite $GdPO_4$, $EuPO_4$, $NdPO_4$, and $LaPO_4$. However, it agrees with what was found for high-pressure monazite-type $CaSO_4$ [29]. This distinctive behavior of high-pressure monazites is related to their more compact structure, which has $AO_9$ polyhedra that are distorted in comparison with low-pressure monazite. The enhancement of this distortion and the increase of the strength of the P-O bonds [6] is what cause the distinctive behavior of the *b*-axis in monazite-type $YPO_4$.

From the pressure dependence of the structural parameters of the low- and high-pressure phases of $YPO_4$ and $ErPO_4$ we have determined the unit-cell volume as a function of pressure. The obtained results are shown in Fig. 7, Clearly, the zircon phase is more compressible than the monazite phase. The fitting of these data to a third-order Birch-Murnaghan EOS [22] gives the following results for $YPO_4$: $V_0$ = 285.6(8) Å$^3$, $B_0$ = 149(2) GPa, and $B_0$' = 3.8(3) for zircon $YPO_4$ and $V_0$ = 265.1(7) Å$^3$, $B_0$ = 206(4) GPa, and $B_0$' = 4.0(2) for monazite $YPO_4$. In the case of $ErPO_4$ we obtained: $V_0$ = 281.5(8) Å$^3$, $B_0$ = 168(4) GPa, and $B_0$' = 4.2(3) for zircon and $V_0$ = 264.5(7) Å$^3$, $B_0$ = 208(6) GPa, and $B_0$' = 4.2(2) for monazite. In contrast with the results of Zhang *et al.* [11], the bulk moduli obtained for the zircon and monazite phase compare well with theoretical estimations [6, 11] (see Table II). Also the increase of this parameter observed after the phase transition is in agreement with the changes observed in other phosphates after a similar collapse of the volume [9].

**D. Bulk modulus and high-pressure systematic of orthophosphates**

We will now discuss the sequence of phases found in different orthophosphates and try to provide a systematic understanding of it, aiming to predict possible phase transitions in unstudied compounds. We have found than the zircon structured compounds $ErPO_4$ and $YPO_4$ transform to monoclinic monazite below 20 GPa. The same transition was found in $YPO_4$ under non-hydrostatic conditions at 15 GPa [11]. A similar tetragonal-monoclinic transition is consistent with the Raman studies performed in isomorphic $TbPO_4$ [10] and $TmPO_4$ [30], which reported phase transitions near 10 and 16

GPa, respectively, and also with *ab initio* calculations recently performed for $TbPO_4$ [31]. However, according to Raman experiments the transition is not reversible, while we and Zhang *et al.* [11] found the transition to be reversible. This discrepancy can be caused by the use of different pressure-transmitting media in the experiments. Whereas we used neon in order to guarantee near hydrostatic conditions in the experiments, the Raman measurements were done using a 4:1 ethanol-methanol mixture, which provides very poor hydrostaticity [17, 24]. It is known, that non-unixial stresses could affect the structural sequence of oxides like the orthophosphates [32].

As we comment above, the zircon-monazite transition is fully expected from the crystal-chemical point of view for those zircons with a large A cation [8]. Indeed, monazite is the low-temperature form of $TbPO_4$ and usually cooling induces a contraction of the crystal structure, which, in the broadest sense, can be considered equivalent to compression. The zircon-monazite transition is a first order transition that involves a collapse in the volume. It also involves a change in polyhedron coordination of the A cation. These atomic rearrangements are accomplished by breaking an A-O bond in zircon and adding two new A-O bonds, which makes monazite much more compact than zircon. Basically, these rearrangements produce the addition of a bond in the equatorial plane of the A cation polyhedron, which is added in the void space between the polyhedra of zircon. These atomic changes occur together with a shift of the (001) planes and a slight rotation of the $PO_4$ tetrahedra, favoring the observed volume collapse. This structural contraction is consistent with the fact that the volume of zircon $TbPO_4$ (291 Å$^3$) is larger than the volume of monazite $GdPO_4$ (279 Å$^3$) despite the fact that the $TbO_8$ polyhedron has a smaller volume (23.7 Å$^3$) than the $GdO_9$ polyhedron (29.4 Å$^3$). Thus, the monazite phase is expected to be less compressible than the zircon phase because the void space in monazite is much smaller.

In contrast with the above-described compounds, zircon-structured orthophosphates with small A cations, undergo a zircon-scheelite transition [9] instead of the zircon-monazite transition. Apparently, upon compression they behave more similar to orthovanadates [23], orthosilicates [33, 34], and orthogermanates [35] than to the rest of the compounds of their own family. The distinctive behavior of $LuPO_4$, $ScPO_4$, and $YbPO_4$ can be related with that fact that in these compounds the ionic radius of the A cations is small relative to that of the $PO_4$ tetrahedra. Therefore, increasing repulsive and steric stresses induced by pressure can be accommodated by significant changes in its average position [36], thereby favoring the reconstructive mechanism involved in the zircon-scheelite transition [37]. The more drastic atomic rearrangement that takes place at the zircon-scheelite transition is what makes this transition irreversible while the zircon-monazite transition is reversible.

In comparison with zircon orthophosphates, monazite phosphates are much more stable under compression. As we have shown, $LaPO_4$ is stable up to 26 GPa, and $GdPO_4$, $EuPO_4$, and $NdPO_4$ up to higher pressures. This result is consistent with the more efficient atomic packing of monazite. The transformation of monazite will be produced only when an increase of the atomic coordination is favored as we found for $LaPO_4$. With the bulk of experimental data available for rare-earth orthophosphates, we have built the qualitative structural systematic shown in Fig. 8. There we also include $YPO_4$. Note that $YPO_4$ follows roughly the same systematic than the rest of the compounds, thus we expect $ScPO_4$ also to follow it. Following section 3.B. we include a possible phase boundary for the zircon-barite transition in monazite phosphates. The transition pressure increases as the atomic radius of the A cation decreases. We also have drawn a phase-boundary for the zircon-monazite transition, which allows us to predict that $DyPO_4$ and $HoPO_4$ will also transform into the monazite structure below 15 GPa. Fig. 8 allows us to make additional predictions. First, $ScPO_4$ (the mineral pretulite) should also transform from the zircon to the scheelite structure beyond 20 GPa, which agree with recent experimental findings [11]. Second, scheelite could be a post-monazite structure for those compounds with small A cations. This suggestion is in good agreement with theoretical calculations made for $TbPO_4$ [29] and $YPO_4$ [11], where a zircon-monazite-scheelite structural sequence is predicted. It also agrees with the systematic proposed by Errandonea and Manjon for $ABO_4$ oxides [8]. The structural systematic shown in Fig. 8 does not include other potential structures, like fergusonite [38], and does not resolve issues like whether compounds like $GdPO_4$ will directly transform from monazite to a barite-related structure or will do it through an intermediate phase (probably scheelite). Further studies are needed to accurately determine the structural sequence of all phosphates; however, Fig. 8 is a practical tool for searching high-pressure transformations in $APO_4$ compounds.

We will discuss now the bulk compressibility of orthophosphates. In order to do it, in Table II, we summarized the bulk modulus of the low- and high-pressure phase of different compounds. The first conclusion we obtain is that zircon and monazite phosphates have a larger bulk modulus than other phosphates of similar stoichiometry in which the phosphorus is in six-fold coordination (e.g. $AlPO_4$ and $FePO_4$) [39]. We can also conclude that for $YPO_4$ the present bulk modulus (149 GPa) agrees better with theory (144 – 165 GPa) and ultrasound measurements (132 GPa) than the bulk modulus obtained under non-hydrostatic conditions (186 GPa). Thus, it is possible that also the non-hydrostatic bulk modulus of monazite $YPO_4$ and low- and high-pressure $ScPO_4$ could be also overestimated. Table II allows also to conclude that the high-pressure phases always have a bulk modulus at least 20 % larger than the low-pressure phases From this table, it is also straightforward to see that within zircon or monazite phosphates, as happen with the vanadates [23], there is an inverse relationship between the atomic volume and the bulk modulus. Consequently, $ScPO_4$ is expected to be the least compressible $APO_4$ compound as found in Ref. [11]. However, it should be mention that the bulk modulus of 376(8) GPa [11] is probably an overestimated value as discussed above. Note that this value is at least 30 % larger than the same parameter in any other scheelite-structured $ABO_4$ oxide. In particular, the reported bulk modulus for scheelite $ScPO_4$ is 70 % larger than that of scheelite $ScVO_4$ [25] and more than 66 % larger than that of the other scheelite-structured phosphates (see Table II). We think the anomalous large bulk modulus reported for scheelite $ScPO_4$ can be affected by non-hydrostatic conditions and the small number of data points collected for this structure [11]. To close this discussion we would like to add that the bulk moduli obtained from chemical bond theory [6] compare pretty well with experiment for the zircons. However, in the case of the monazites, this theory underestimates by more than 10 % the value of the bulk modulus.

For the ambient pressure phase of $ABO_4$ compounds that crystallize in the scheelite or zircon structures, the bulk modulus can be directly correlated to the compressibility of the $AO_8$ polyhedron. For most of these compounds the bulk modulus obeys the following empirical formula [43]: $B_0 = 610\ Z_A/d^3_{A\text{-}O}$; where $B_0$ is the bulk modulus in GPa, $Z_A$ is the formal charge of the A-cation, and $d_{A\text{-}O}$ is the average A-O distance (in Å) in the $AO_8$ polyhedron at ambient pressure. In Table II it can be seen than the estimates obtained using this empirical formula are as good as the theoretical calculated values for zircons. Therefore, it can be used as a first approximation to determine the bulk modulus of compounds like $TmPO_4$ (146 GPa) and $HoPO_4$ (142 GPa). If we apply the same empirical formula to monazites, we find that it underestimates the bulk modulus, even more than theoretical calculations do. A possible reason for it is related to the structural differences between zircon and monazite. Remember that if we compare both structures, we find that the inter-polyhedral empty space of zircon tend to be filled in monazite by a new A-O bond. As a consequence of it, the $AO_9$ polyhedra of monazite are distorted and more densely packed, being some of the A-O bonds less compressible in monazite than in zircon. To check this hypothesis, we extracted from the experimental data the bond distances of monazite $LaPO_4$ and zircon $YPO_4$ as a function of pressure. The results are summarized in Figs. 9 and 10. There it can be seen that, in zircon the A-O bonds, are much more compressible than the P-O bonds. Consequently they account for most of the volume reduction, and the empirical relation can be applied. In the case of monazite, we have a more complicated scenario. Three P-O bonds are rigid, but the remaining one is more compressible than the others. In the case of the A-O bonds of monazite, we have four rigid bounds and five compressible bonds. The more rigid bonds are those with the longest projection along the *c*-axis and the most compressible bonds are aligned along the *a-b* plane. Therefore, it is clear that the bulk compressibility of monazite phosphates cannot be explained within the framework developed for zircons and scheelites [8].

Basically, since only some of the A-O bonds are highly compressible in monazite, the empirical relation should underestimate the bulk modulus of monazites. This is exactly what we found as can be seen in Table II.

### 4. Concluding remarks

We performed RT ADXRD measurements on $LaPO_4$, $NdPO_4$, $EuPO_4$, $GdPO_4$, $ErPO_4$, and $YPO_4$ up to pressures close to 30 GPa using neon as pressure-transmitting medium. In $LaPO_4$ we found the onset of a phase transition from monazite to a more symmetric structure at near 26 GPa. For the high-pressure phase we proposed a barite-type structure and apparently the phase transition is non-reversible. In $NdPO_4$, $EuPO_4$, and $GdPO_4$ we found that the monazite structure remains stable up to 30 GPa. In $YPO_4$ and $ErPO_4$ we detected a phase transition from zircon to monazite at 20 GPa and 17 GPa, respectively. The transition is reversible upon decompression. The reported phase transformations are consistent with the structural sequence deduced using crystal-chemistry arguments from other $ABO_4$ oxides [8]. In addition, based upon the present and previous results a structural systematic for orthophosphates is proposed. From the experiments, we also obtained the axial and bulk compressibility of the different compounds. We found that compression is anisotropic and determined the equation of state for the different phases. In particular, $ScPO_4$ is proposed to be the less compressible zircon-structured orthophosphate. We also found that in a given compound the monazite structure is less compressible than the zircon structure due to the large packing efficiency of monazite in comparison to that of zircon. Finally, for zircon $YPO_4$ we found a differential polyhedral compressibility, being the $PO_4$ tetrahedra much stiffer than the $YO_8$ dodecahedra. In the case of monazite $LaPO_4$ we found a different behavior. In this material not only the P-O bonds but also some of the A-O bonds show an uncompressible nature. These facts have been related with the anisotropic compressibility of both structures.

**Acknowledgments**

Financial support from Spanish MALTA-Consolider Ingenio 2010 Program (Project CSD2007-00045) is gratefully acknowledged. This work was partially supported by the Spanish MICCIN under Grant No. MAT2007-65990-C03-01. This work was performed at HPCAT (Sector 16), Advanced Photon Source (APS), Argonne National Laboratory. HPCAT is supported by DOE-BES, DOENNSA, NSF, and the W.M. Keck Foundation. APS is supported by DOE-BES, under Contract No. DE-AC02-06CH11357. The authors thank Dr. Stanislav Sinogeikin of HPCAT and Erica Viviani of Univ. Verona for technical support. R. L.-P. thanks the support of the MICINN through the FPU program.

## References

[1] Y. Ni, J. M. Hughes, and A. N. Mariano, Amer. Mineral. **80**, 21 (1995).

[2] A. Meldrum, L. A. Boatner, and R.C. Ewing, Phys. Rev. B **56**, 13805 (1997) and references therein.

[3] D. Rubatto, J. Hermann, and I. S. Buick, J. Petrology **47**, 1973 (2006).

[4] U. Kolitsch and D. Holtstam, Eur. J. Mineral. **16**, 117 (2004).

[5] A. A. Kaminskii, M. Bettinelli, A. Speghini, H. Rhee, H. J. Eichler, G. Mariotto, Laser Phys. Lett. **5**, 367 (2008).

[6] H. Li, S. Zhang, S. Zhou, and X. Cao, Inorganic Chemistry **48**, 4542 (2009).

[7] R. S. Hay, Ceram. Eng. Sci. Proc. **21**, 203 (2001).

[8] D. Errandonea and F. J. Manjon, Progress in Materials Science **53**, 711 (2008).

[9] F. X. Zhang, M. Lang, R. C. Ewing, J. Lian, Z. W. Wang, J. Hu, and L. A. Boatner, J. Solid State Chem. **181**, 2633 (2008).

[10] A. Tatsi, E. Stavrou, Y. C. Boulmetis, A. G. Kontos, Y. S. Raptis, and C. Raptis, J. Phys.: Condens. Matter **20**, 425216 (2008).

[11]F. X. Zhang, J. W. Wang, M. Lang, J. M. Zhang, and R. C. Ewing, Phys. Rev. B **80**, 184114 (2009).

[12] G. Chen, J. Holsa, and J. R. Peterson, J. Phys. Chem. Solids **58**, 2031 (1997).

[13] R. S. Feigelson, J. Am. Ceram. Soc. **47**, 257 (1964).

[14] H. K. Mao, J. Xu, and P. M. Bell, J. Geophys. Res. **91**, 4673 (1986).

[15] A. Dewaele, P. Loubeyre, and M. Mezouar, Phys. Rev. B **70**, 094112 (2004).

[16] A. Dewaele, F. Datchi, P. Loubeyre, and M. Mezouar, Phys. Rev. B **77**, 094106 (2008).

[17] S. Klotz, J. C. Chervin, P. Munsch, and G. Le Marchand, J. Phys. D: Appl. Phys. **42**, 075413 (2009).

[18] A. P. Hammersley, S. O. Svensson, M. Hanfland, A. N. Fitch, and D. Häusermann, High Press. Res. **14**, 235 (1996).

[19] W. Kraus and G. Nolze, J. Appl. Crystallogr. **29**, 301 (1996).

[20] A. C. Larson and R. B. Von Dreele, LANL Report 86–748 (2000).

[21] W. A. Crichton, J. B. Parise, S. M. Antao, amd A. Grzechnik, Amer. Minerl. **90**, 22 (2005).

[22] F. Birch, J. Geophys. Res. **83**, 1257 (1978).

[23] O. Tschauner, P. Dera, and B. Lavina, Acta Cryst. A **64**, C608 (2008).

[24] D. Errandonea, Y. Meng, M. Somayazulu, and D. Häusermann, Physica B **355**, 116 (2005).

[25] D. Errandonea, R. Lacomba-Perales, J. Ruiz-Fuertes, A. Segura, S. N. Achary, and A. K. Tyagi, Phys. Rev. B **79**, 184104 (2009).

[26] D. Errandonea, R. S. Kumar, X. Ma, and C. Tu, J. Sol. State Chem. **181**, 355 (2008).

[27] L. A. Boatner and B. C. Sales, *Radioactive waste forms for the future*, edited by W. Lutze and R. C. Ewing (Elsevier, Amsterdam, 1988), pp. 495 –564.

[28] H. Nyman, B. G. Hyde, and S. Andersson, Acta Cryst. B **40**, 441 (1984).

[29] S. E. Bradbury and Q. Williams, J. Phys. Chem. Solids **70**, 134 (2009).

[30] E. Stavrou, C. Raptis, and K. Syassen, Proceedings of the 47th EHPRG Conference (Paris 2009).

[31] J. Lopez-Solano, P. Rodriguez-Hernandez, and A. Muñoz, Proceedings of the 47th EHPRG Conference (Paris 2009).

[32] R. Lacomba-Perales, D. Martínez-García, D. Errandonea, Y. Le Godec, J. Philippe, and G. Morard, High Pressure Research **29**, 76 (2009).

[33] H. P. Scott, Q. Williams, and E. Knittle, Phys. Rev. Letters **88**, 015506 (2002).

[34] L. Gracia, A. Beltrán, and D. Errandonea, Phys. Rev. B **80**, 094105 (2009).

[35] D. Errandonea, R. S. Kumar, L. Gracia, A. Beltrán, S. N. Achary, and A. K. Tyagi, Phys. Rev. B **80**, 094101 (2009).

[36] D. Errandonea, N. Somayazulu, and D. Häusemann, phys. stat. sol. (b) **235**, 162 (2003).

[37] M. Florez, J. Contreras-Garcia, J. M. Recio, and M. Marques, Phys. Rev. B **79**, 104101 (2009).

[38] D. Errandonea, phys. stat. sol. (b) **242**, R125 (2005).

[39] S. M. Sharma, N. Garg, and S. K. Sikka, Phys. Rev. B **62**, 8824 (2000).

[40] A. Ambruster, J. Phys. Chem. Solids **37**, 321 (1976).

[41] P. Mogilevsky, E. Zaretsky, T. Parthasarathy, and F. Meisenkothen, Phys. Chem. Miner. **33**, 691 (2006).

[42] J. Wang, Y. Zhou, and Z. Lin, Appl. Phys. Letters **87**, 051902 (2005).

[43] D. Errandonea, J. Pellicer-Porres, F. J. Manjon, A. Segura, Ch. Ferrer-Roca, R. S. Kumar, O. Tschauner, P. Rodriguez-Hernandez, J. Lopez-Solano, S. Radescu, A. Mujica, A. Muñoz, and G. Aquilanti, Phys. Rev. B **72**, 174106 (2005).

**Table I:** Unit-cell parameters as a function of pressure for the ambient-pressure phases of monazite and zircon orthophosphates. Pressure in GPa, *a*, *b*, and *c* in Å, and β in degrees.

| | **$GdPO_4$** | **$EuPO_4$** | **$NdPO_4$** |
|---|---|---|---|
| *a(P)* | 6.623(6) - 0.0120(3) *P* | 6.613(6) - 0.0123(3) *P* | 6.706(4) - 0.0127(3) *P* |
| *b(P)* | 6.829(7) - 0.0104(4) *P* | 6.861(7) - 0.0108(4) *P* | 6.925(5) - 0.0098(3) *P* |
| *c(P)* | 6.335(8) - 0.0089(4) *P* | 6.349(8) - 0.0091(4) *P* | 6.392(6) - 0.0083(3) *P* |
| *β(P)* | 103.80(6) –0.051(3) *P* | 103.90(6) -0.055(3) *P* | 103.55(6) - 0.063(4) *P* |

| | **$LaPO_4$** | **$YPO_4$** | **$ErPO_4$** |
|---|---|---|---|
| *a(P)* | 6.808(4) - 0.0160(3) *P* | 6.877(6) - 0.0146(3) *P* | 6.864(5) - 0.0144(4) *P* |
| *b(P)* | 7.061(4) - 0.0127(3) *P* | | |
| *c(P)* | 6.478(7) - 0.0081(4) *P* | 6.017(7) - 0.0071(8) *P* | 5.999(4) - 0.0060(6) *P* |
| *β(P)* | 103.28(3) – 0.092(2) *P* | | |

**Table II:** Bulk modulus (given in GPa) and unit-cell volume at ambient pressure (in $Å^3$) of different orthophosphates. Volumes are given only for those structures that are stable at ambient pressure. Experimental and theoretical results are included for $B_0$. Note that different pressure media were employed in different experiments (see Refs.).

| Compound | Structure | Unit-cell volume | Bulk Modulus | | |
|---|---|---|---|---|---|
| | | | Experiments | Theory | Empirical model |
| $ScPO_4$ | zircon | 252.1 | 203(7)[a] | 175.1[f] – 183[a] | 169 |
| $ScPO_4$ | zircon | | 376(8)[a] | 334[a] | |
| $LuPO_4$ | zircon | 273.7 | 184(4)[b] – 166[c] | 152.8[f] | 150 |
| $LuPO_4$ | scheelite | | 226(3)[b] | | |
| $YbPO_4$ | zircon | 276.5 | 150(5)[b] | 150[f] | 147 |
| $YbPO_4$ | scheelite | | 218(2)[b] | | |
| $TmPO_4$ | zircon | 278.9 | | 147.2[f] | 146 |
| $ErPO_4$ | zircon | 281.5 | 168(4)[e] | 146.1[f] | 145 |
| $ErPO_4$ | monazite | | 208(6)[e] | | |
| $YPO_4$ | zircon | 286.5 | 132[d] - 149(2)[e] – 186(5)[a] | 144.4[f] – 165[a] | 143 |
| $YPO_4$ | monazite | | 206(4)[e] – 260[a] | 190[a] | |
| $YPO_4$ | scheelite | | | 213.7[a] | |
| $HoPO_4$ | zircon | 284.6 | | 143.4[f] | 142 |
| $DyPO_4$ | zircon | 287.9 | | 141.5[f] | 141 |
| $TbPO_4$ | zircon | 291.4 | | 138.8[f]-128[h] | 139 |
| $GdPO_4$ | monazite | 279.1 | 160(2)[e] | 149[f] | 120 |
| $EuPO_4$ | monazite | 281.6 | 159(2)[e] | 147.1[f] | 118 |
| $SmPO_4$ | monazite | 284.4 | | 146[f] | 117 |
| $NdPO_4$ | monazite | 291.4 | 170(2)[e] | 142.3[f] | 114 |
| $PrPO_4$ | monazite | 295.3 | | 139.7[f] | 112 |
| $CePO_4$ | monazite | 299.5 | | 137.2[f] | 110 |
| $LaPO_4$ | monazite | 305.7 | 144(2)[e] | 134[f] – 100[g] | 107 |
| $LaPO_4$ | barite | 296.2 | 143(4)[e] | | |

[a] Ref 11 [b]Ref. 9, [c]Ref. 40, [d]Ref. 41, [e]Present study, [f]Ref. 6, [g]Ref. 42, [h]Ref. 31.

**Figure Captions**

**Figure 1 (colour online):** Schematic view of the zircon (left) and monazite (right) structures. Large circles: A cation, middle circles: P, and small circles: O. The $AO_8$ ($AO_9$) and $PO_4$ polyhedra are shown in blue (dark) and yellow (light).

**Figure 2:** Selected x-ray diffraction patterns of $LaPO_4$ at different pressures; (r) indicates the pattern collected on pressure release. The inset show the low-angle section of the spectrum collected at 26.1 GPa enlarged. The peaks appearing at low angles can be more clearly seen there.

**Figure 3:** Unit-cell parameters versus pressure for $LaPO_4$. Circles: low-pressure phase. Triangles: high-pressure phase. The empty symbols correspond to data obtained after pressure release. The dashed lines are a guide to the eye and the solid lines represent the fits shown in Table I. The inset shows the pressure evolution of the $\beta$ angle of the monazite structure.

**Figure 4:** Pressure-volume relation in $LaPO_4$, $NdPO_4$, $EuPO_4$, and $GdPO_4$. Solid symbols: experiments. Empty circles: data points collected after pressure release. Empty square: literature data [1] Lines: EOS fit.

**Figure 5:** Selected x-ray diffraction patterns of $YPO_4$ at different pressures; (r) indicates the pattern collected on pressure release.

**Figure 6:** Unit-cell parameters versus pressure for $YPO_4$. Squares: low-pressure phase. Circles: high-pressure phase. The dashed lines are a guide to the eye and the solid lines represent the fits shown in Table I. The inset shows the pressure evolution of the $\beta$ angle of the monazite structure.

**Figures 7:** Pressure-volume relation in $YPO_4$ and $ErPO_4$. Solid squares: experiments for the zircon phase. Solid circles: data points obtained for the high-pressure phase. Empty triangles: literature data [1]. Empty squares: data obtained after pressure release. Lines: EOS fit. In $YPO_4$ the empty square overlap the empty triangle.

**Figure 8:** Proposed qualitative structural systematic of orthophosphates under pressure at ambient temperature. The compounds are ordered according with the ionic radii of the A cation, going from the smaller cation in the left to the largest cation in the right. Dashed lines represent possible phase boundaries.

**Figure 9:** Pressure dependence of bond distances in monazite $LaPO_4$. Symbols: experiments. Lines: fits.

**Figure 10:** Pressure dependence of bond distances in zircon $YPO_4$. Symbols: experiments. Lines: fits.

**Figure 1**

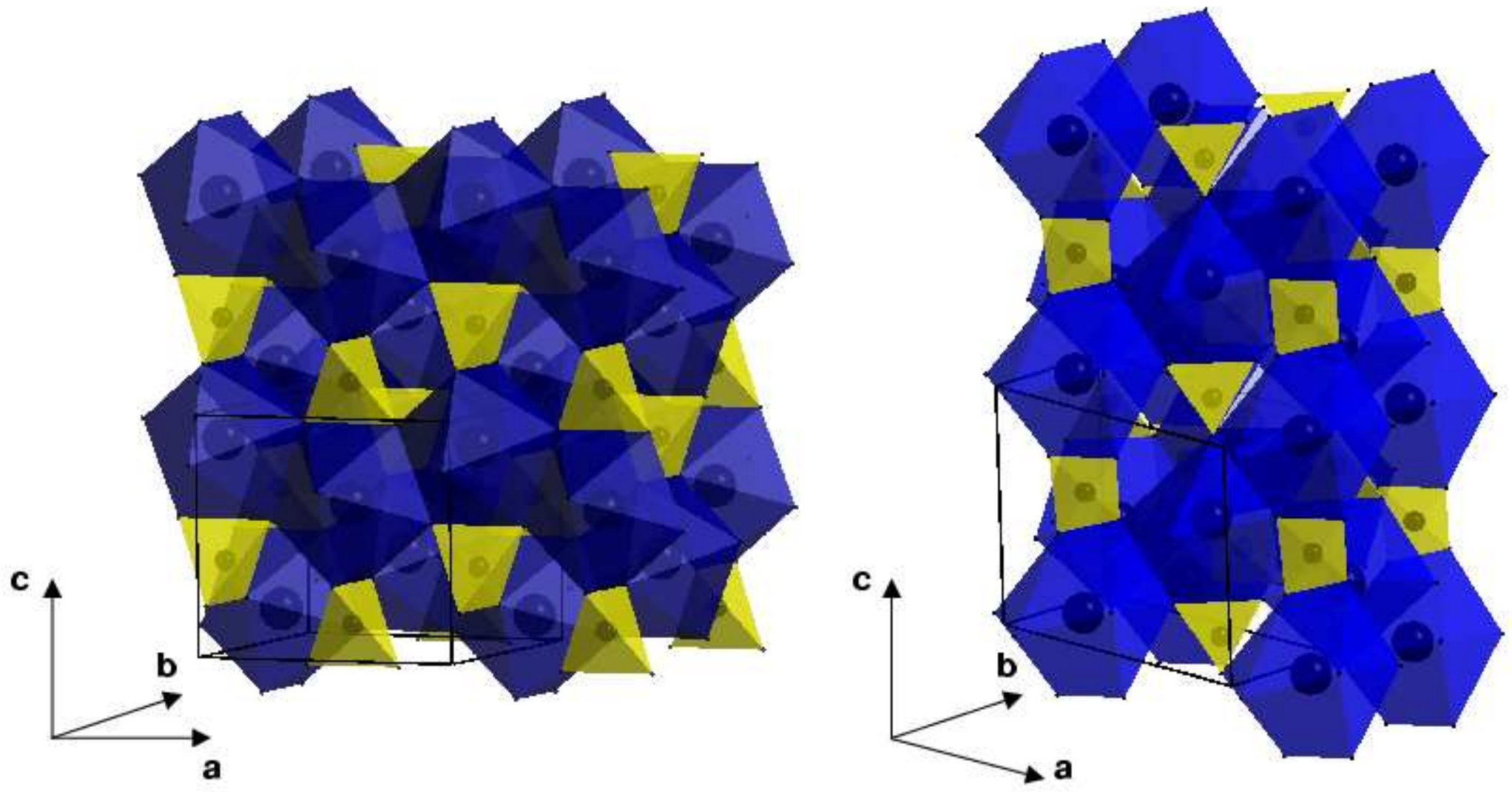

**Figure 2**

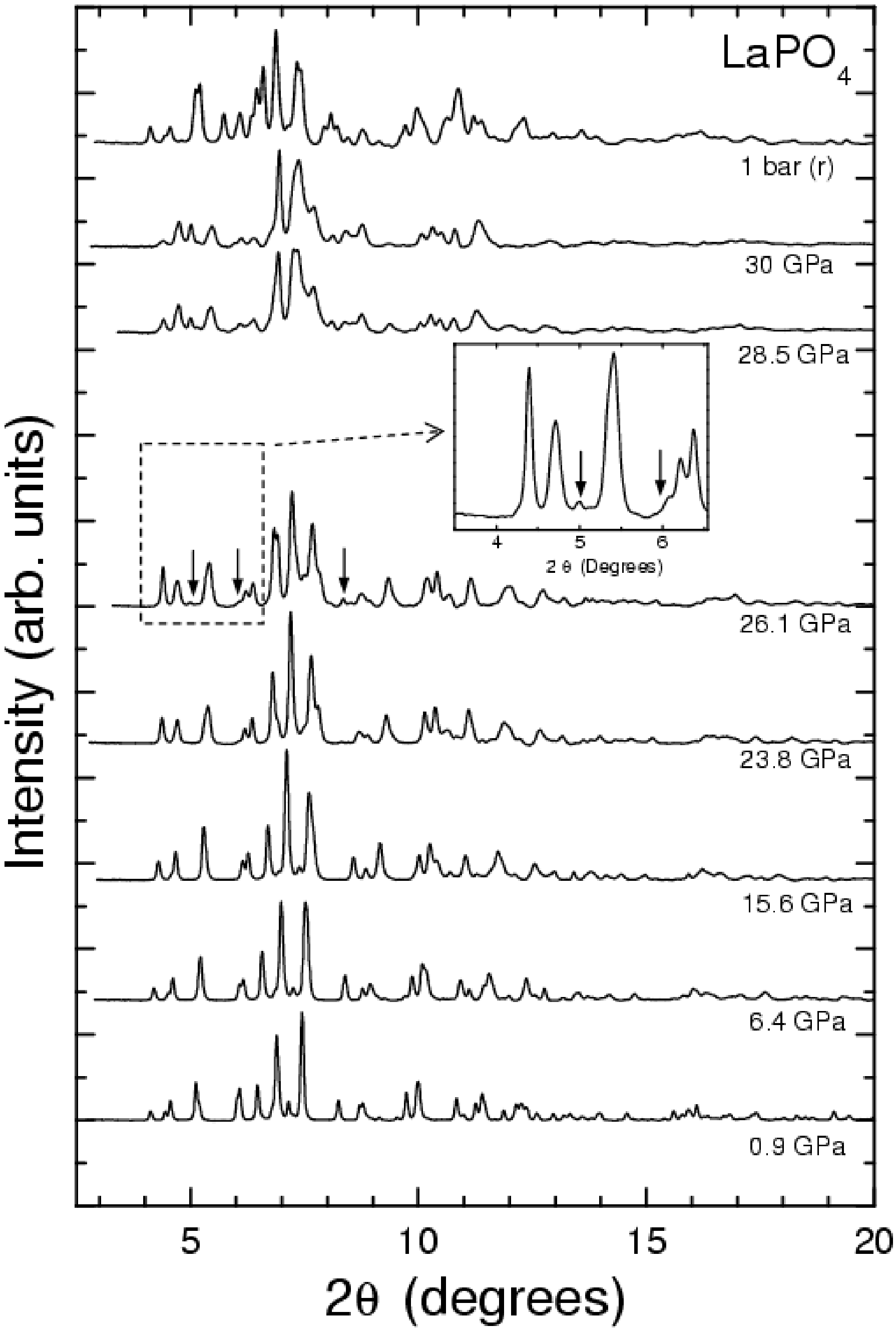

**Figure 3**

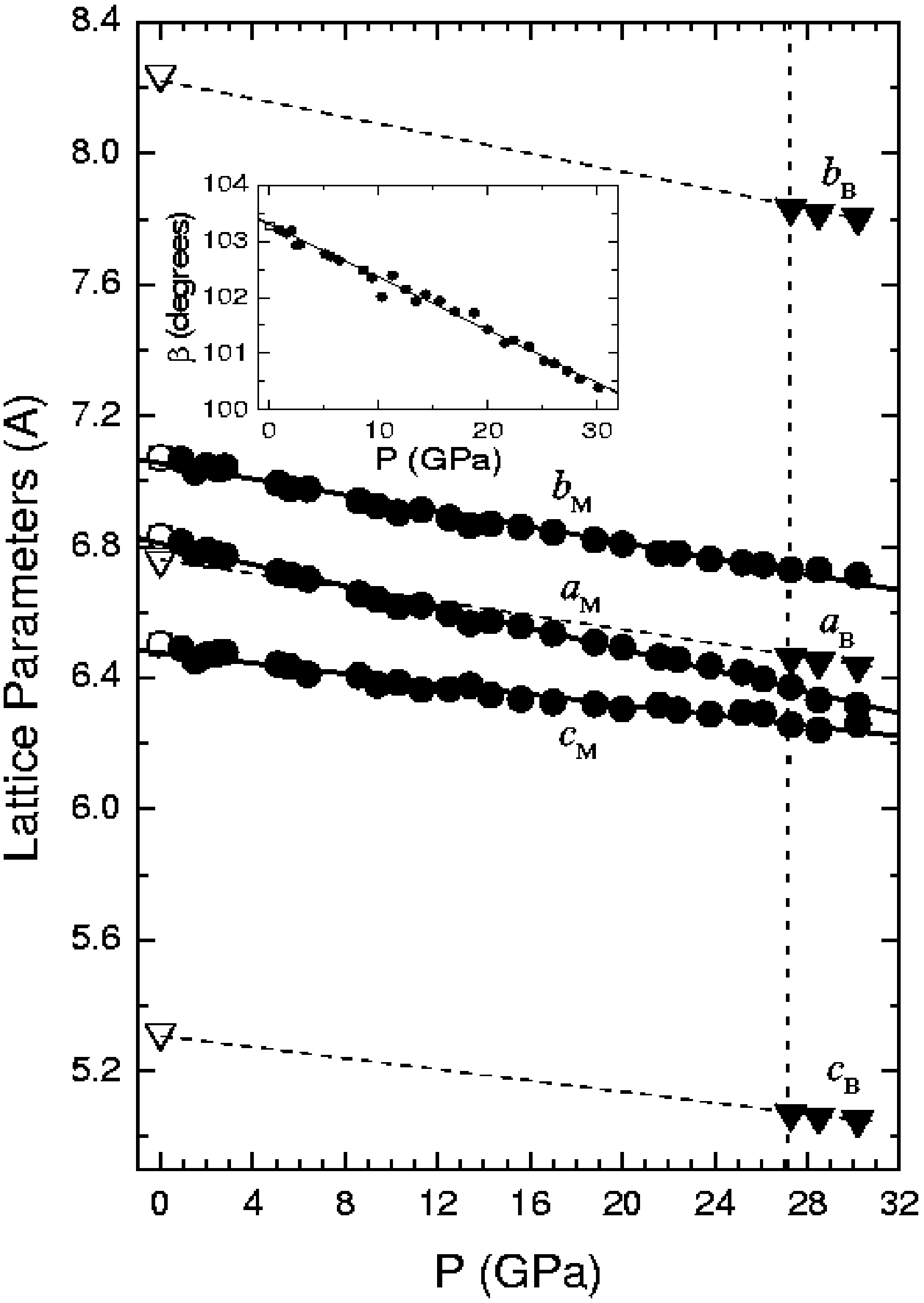

**Figure 4**

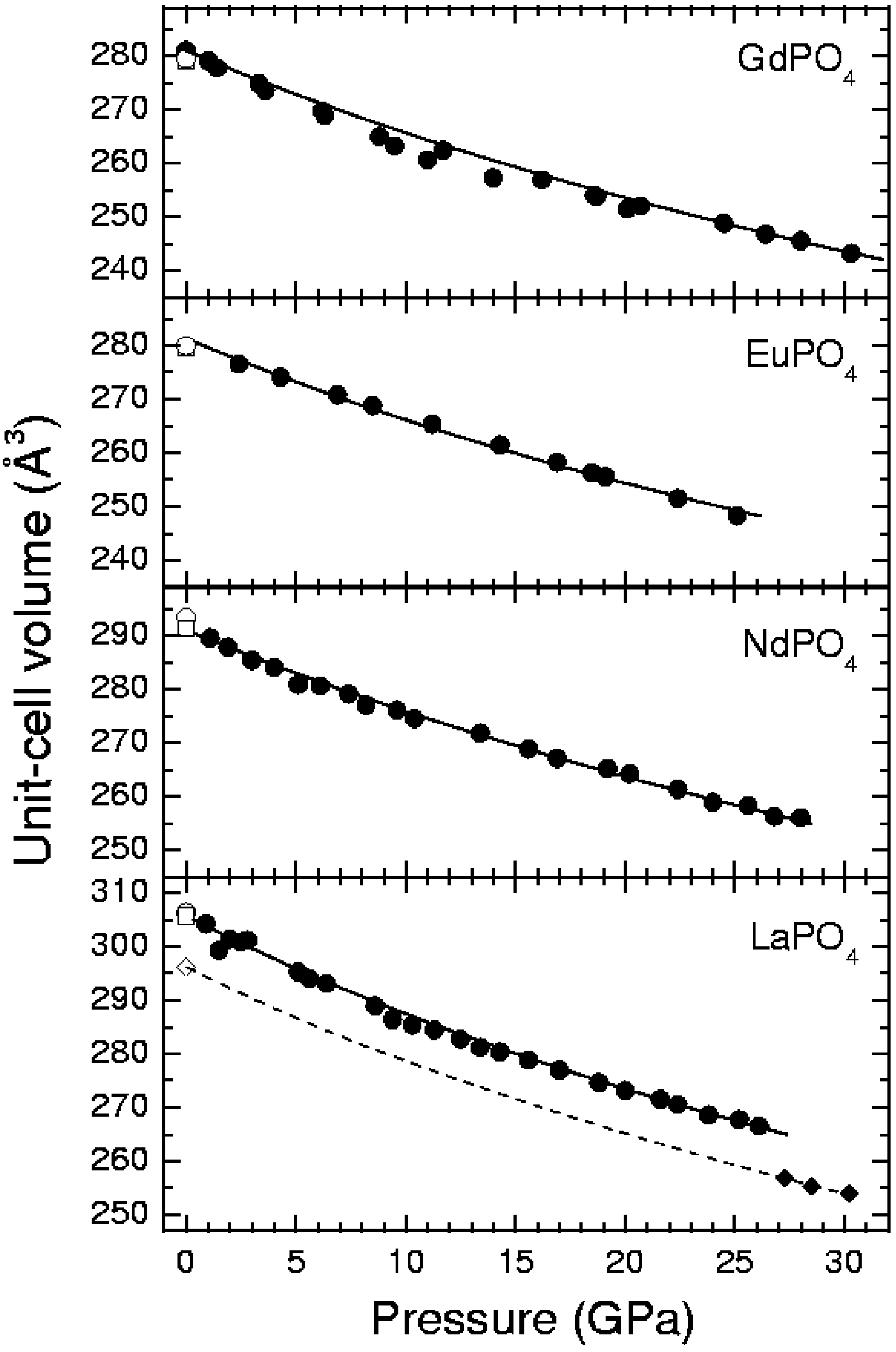

**Figure 5**

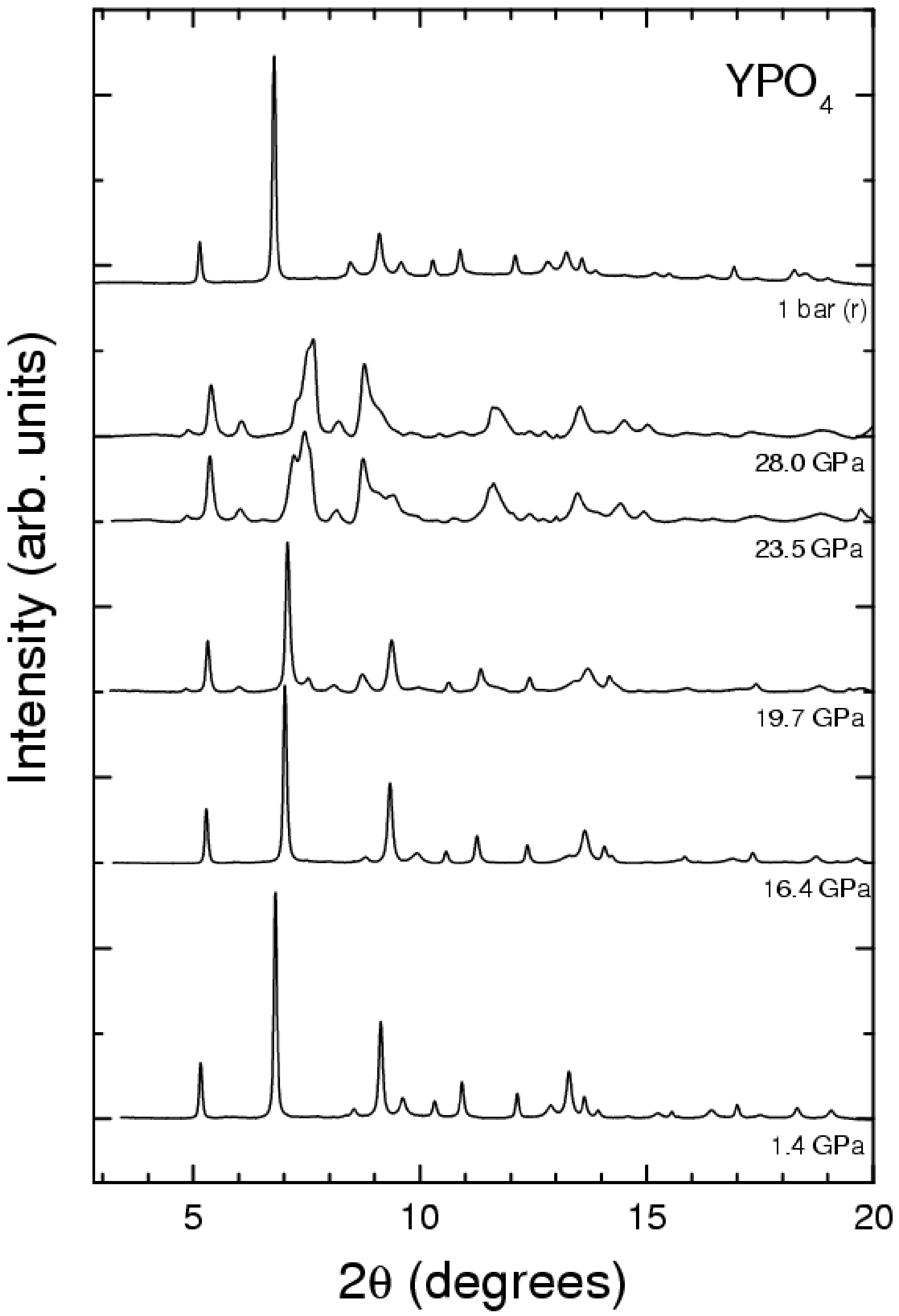

**Figure 6**

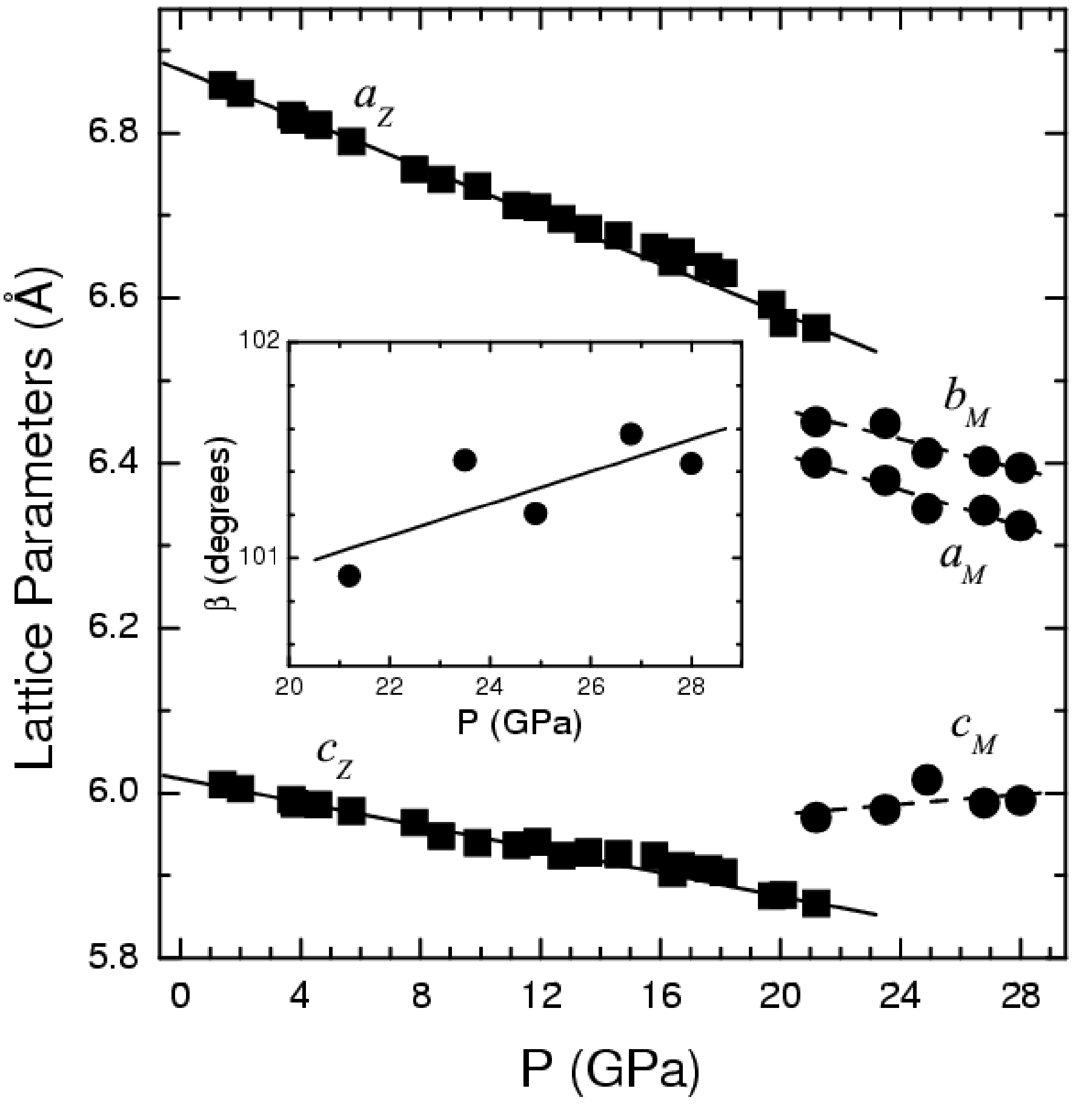

**Figure 7**

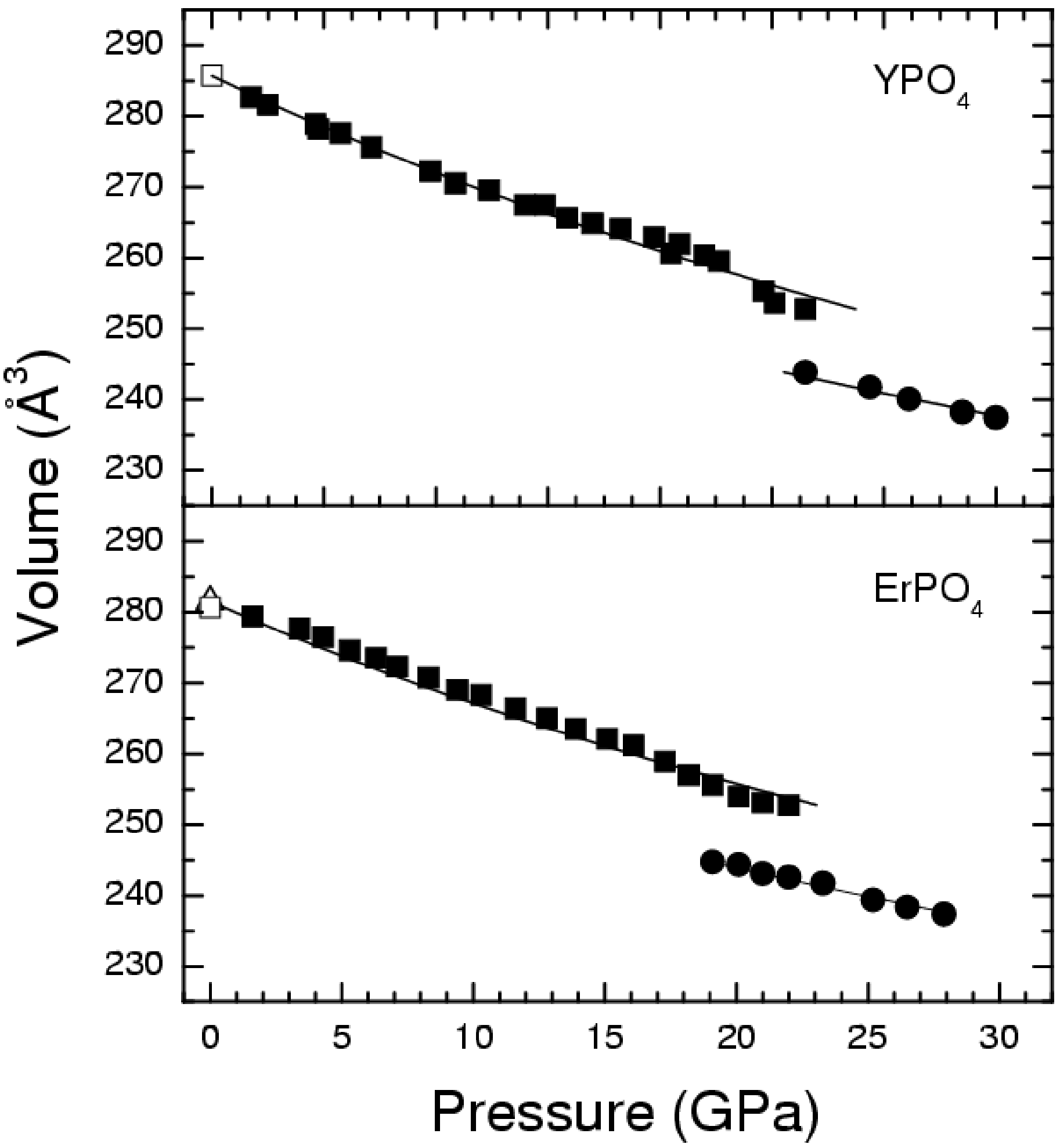

**Figure 8**

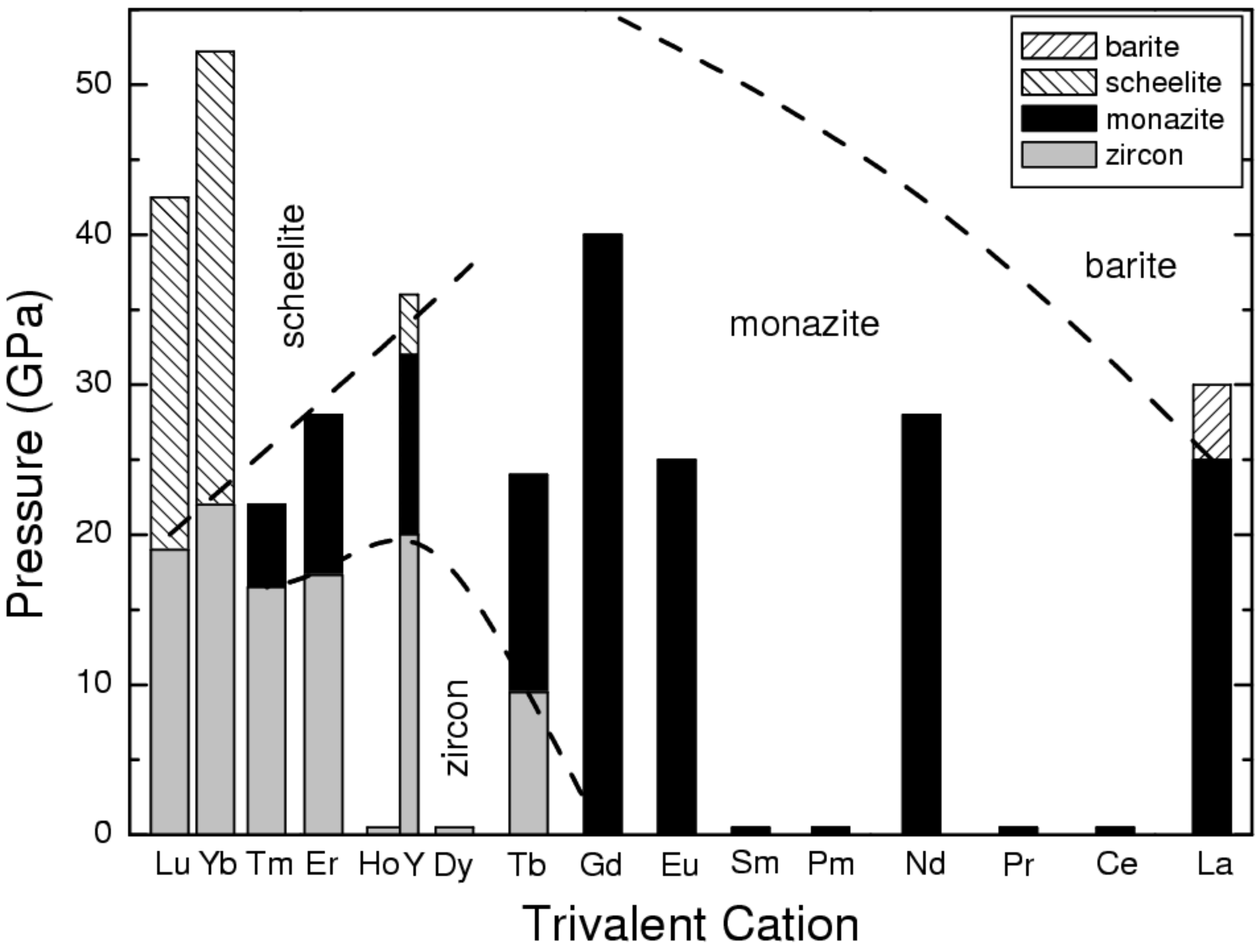

**Figure 9**

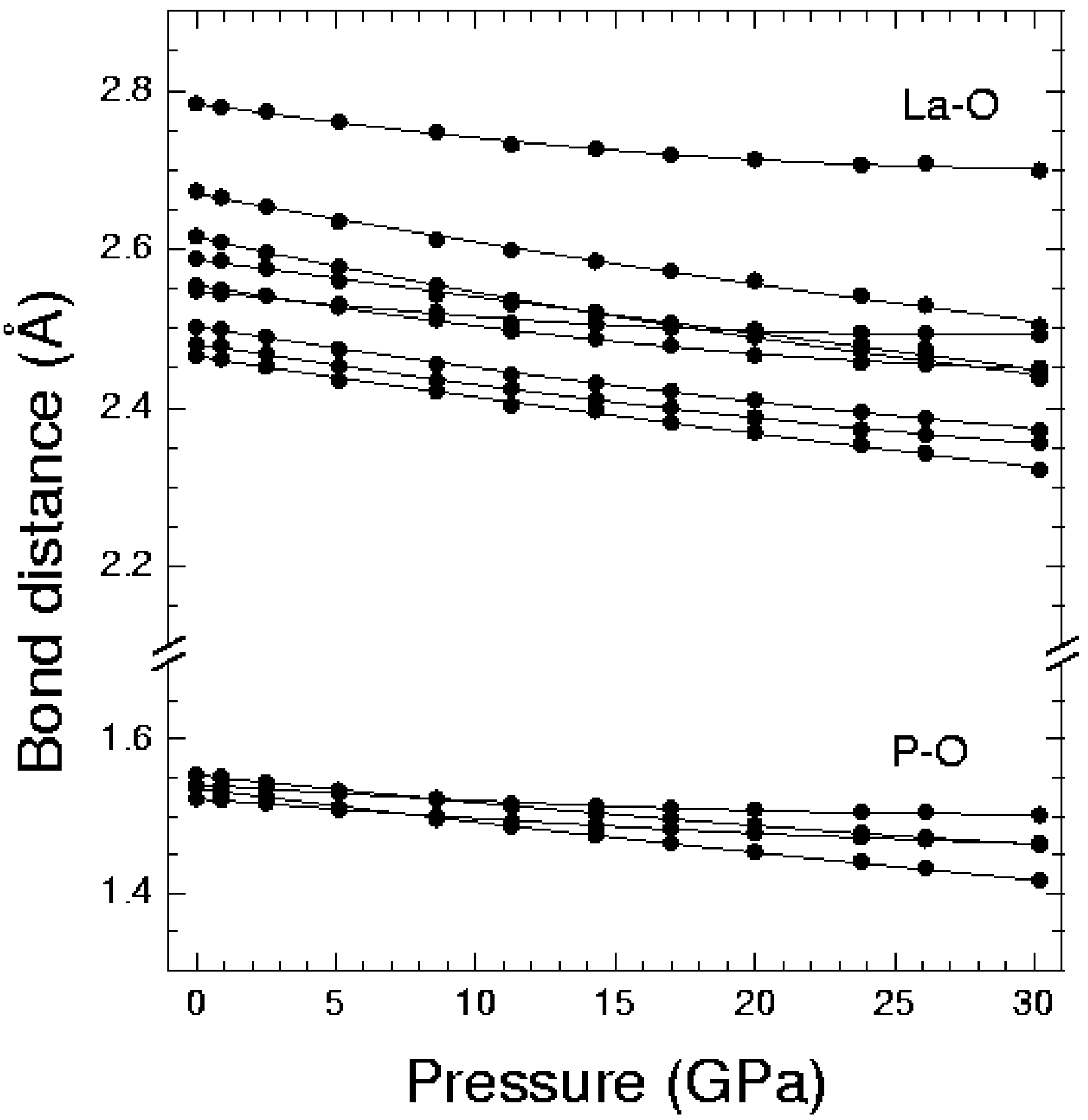

**Figure 10**

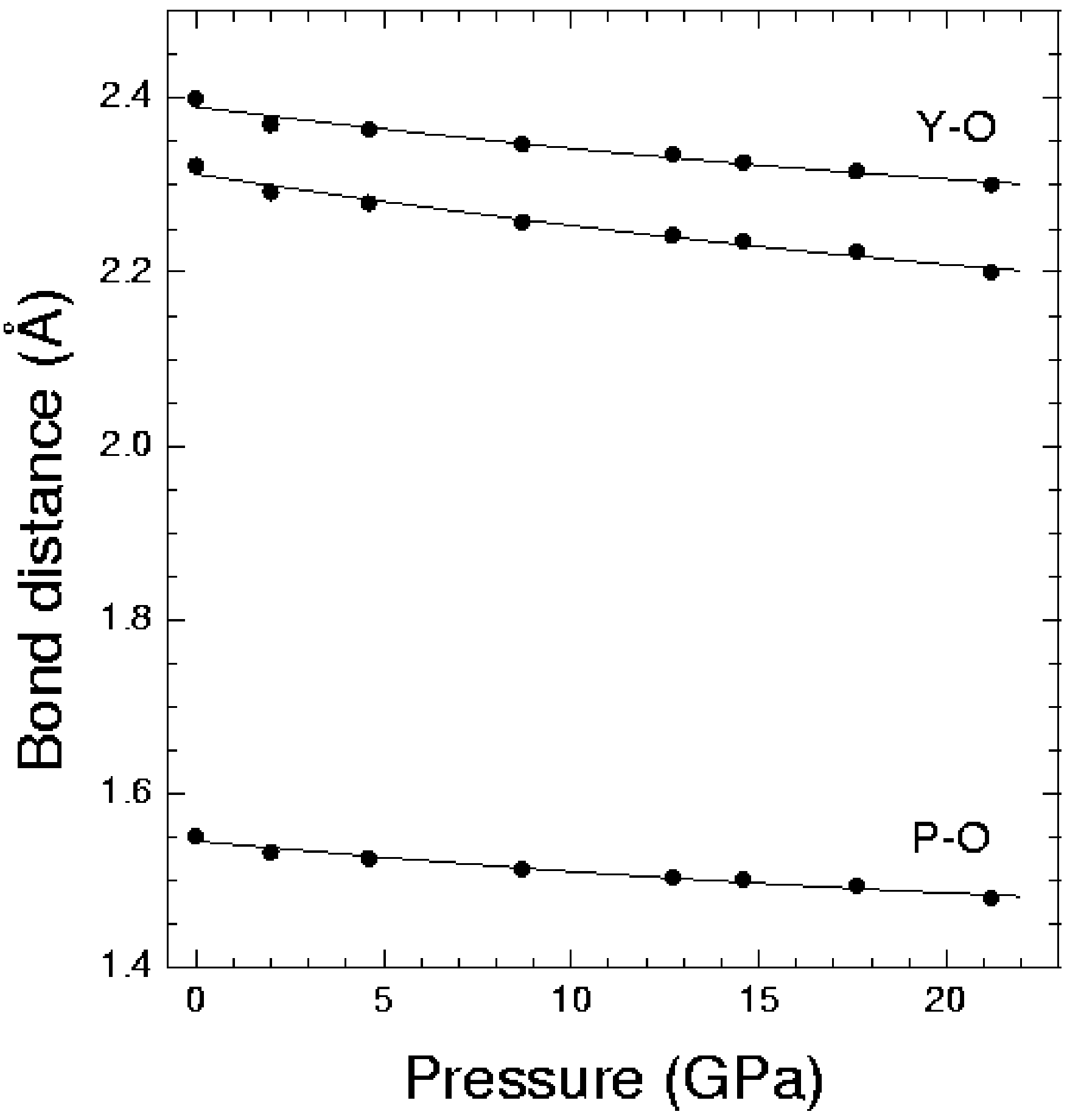